\documentclass[iop,apj,numberedappendix,twocolappendix,revtex4]{emulateapj}

\usepackage{multirow}

\slugcomment{}

\begin{document}
\title{On the stability of a galactic disk in modified gravity}
\shorttitle{On the stability of a galactic disk in modified gravity}
\author{Mahmood Roshan $^{1}$ and Shahram Abbassi $^{1,2}$}
\affil{$^1$Department of Physics, Ferdowsi University of Mashhad, P.O. Box 1436, Mashhad, Iran; mroshan@um.ac.ir, abbassi@um.ac.ir}
\affil{$^2$School of Astronomy, Institute for Research in Fundamental Sciences (IPM), P.O. Box 19395-5531, Tehran, Iran}

\begin{abstract}
We find the dispersion relation for tightly wound spiral density waves in the surface of rotating, self-gravitating disks in the framework of Modified Gravity (MOG). Also, the Toomre-like stability criterion for differentially rotating disks has been derived for both fluid and stellar disks. More specifically, the stability criterion can be expressed in terms of a matter density threshold over which the instability occurs. In other words the local stability criterion can be written as $\Sigma_0<\Sigma_{\text{crit}}(v_s,\kappa,\alpha,\mu_0)$, where $\Sigma_{\text{crit}}$ is a function of $v_s$ (sound speed), $\kappa$ (epicycle frequency) and $\alpha$ and $\mu_0$ are the free parameters of the theory. In the case of a stellar disk the radial velocity dispersion $\sigma_r$ appears in $\Sigma_{\text{crit}}$ instead of $v_s$. We find the exact form of the function $\Sigma_{\text{crit}}$ for both stellar and fluid self-gravitating disks. Also, we use a sub-sample of THINGS catalog of spiral galaxies in order to compare the local stability criteria. In this perspective, we have compared MOG with Newtonian gravity and investigated the possible and detectable differences between these theories.
\end{abstract}
\section{Introduction}
\label{introduction}

Disk galaxies encompass non-axisymmetric features such as spiral arms and bars which have significant influences on the galactic evolution (Sellwood 2014, Kormendy \& Kennicutt 2004). These spiral arms trigger star formation in the outer part of the disk galaxy, drive secular changes in the stellar orbits and change the mass distribution within the disk. Understanding the origin and evolution of these spiral structures is one of the hardest problems in astrophysics. However, it is widely believed that spiral patterns are gravitationally driven  density waves (e.g., those produced by a spontaneous disturbance or, in some cases, a companion system) and propagating in the surface of the galaxies. Therefor the density wave theory is a indispensable tool in this subject. Assuming that density waves are small perturbations on the density of the galactic disk, one can use linear analysis in order to find the linear oscillatory modes of the disk. However, analyzing the behavior of a density wave in a galactic disk and calculating the relevant oscillatory modes is a difficult task. The main reason for this difficulty is that gravity plays the dominant role in the evolution of the density waves. On the other hand, gravity is a long range force. Correspondingly, perturbations in all parts of the system are coupled. This coupling makes the analysis very difficult and consequently the oscillatory modes can be found analytically only for Kalnajs disks (Kalnajs 1972). Also numerical mode calculations have been performed for some simple models including the isochrone disk, the Kuzmin disk, and power-law disks (Zang 1976; Kalnajs 1978; Vauterin \&  Dejonghe 1996; Pichon \&  Cannon 1997; Evans \&  Read 1998; Jalali \& Hunter 2005).

In the early 1960s it was understood that for tightly wound spiral density waves (waves whose wavelength is much less than the radius), the long-range coupling of gravity is negligible, and the response of the background disk to this tightly wound spiral density waves can be determined locally (Lin \& Shu 1964). In fact, in this approximation, known as WKB approximation, the relevant analysis is completely analytic and independent of the given model for the background disk. In other words, using this approximation, one can easily determine the stability of the given disk against axisymmetric perturbations (density wave). At an arbitrary location in the stellar disk, stability requires that
\begin{equation}
Q_s=\frac{\kappa \sigma_r}{3.36 G\Sigma_0 }>1
\label{Q1}
\end{equation}
where $\kappa$ is the epicyclic frequency, $\sigma_r$ is the radial velocity dispersion, $\Sigma_0$ is the disk unperturbed density, and $G$ is the gravitational constant. The requirement (\ref{Q1}) is known as the Toomre's local stability criterion and has a simple physical interpretation. In fact this criterion represents a competition between the stabilizing influences of pressure ($\sigma_r$) and the angular momentum ($\kappa$) against the destabilizing effect of gravity ($\Sigma_0$). In the case of a fluid (gaseous) disk, this criterion takes the following form
\begin{equation}
Q_g=\frac{\kappa v_s}{\pi G\Sigma_0 }>1
\label{Q2}
\end{equation}
where $v_s$ is the sound speed. It is worth mentioning that it was first shown by Safronov (1960) that a thin fluid rotating disc can be unstable against local axisymmetric disturbances under the effect of its own gravity. These criteria, i.e. (\ref{Q1}) and (\ref{Q2}), have been modified for taking into account various effects. For example the effect of the thickness of the disk has been investigated by Toomre (1964) and Vandervoort (1970). Two component (fluid + stars) disks of finite thickness has been considered by Kato (1972), Bertin \& Romeo (1988), Romeo (1992), Wang \& Silk (1994) and Romeo \& Wiegert (2011). See also Shadmehri et al. (2012) for a two component study where the gaseous component is turbulent. Rafikov (2001) studied the general case of multiple components. The stability of a two-fluid disk was investigated by Jog \& Solmon (1984), Elmegreen (1995) and Jog (1996). Effects of magnetic field on linear gravitational instabilities in thin galactic disks have been studied by Elmegreen (1987, 1994), Gammie (1996), Fan \& Lou (1997), Kim and Ostriker (2001). Gammie (1996) has also considered the effect of viscosity to the local stability. Also the local stability in two-component galactic disk with gas dissipation has been studied by Elmegreen (2011).

The purpose of this paper is to introduce a modified stability criterion in the context of Modified gravity (MOG) for a single component galactic disk (fluid or stellar). This theory is a covariant modification of Einstein's general relativity (GR) and has been introduced for addressing the dark matter problem (see Moffat 2006). In the weak field limit, the Poisson equation in this theory is different from that of Newtonian gravity. Therefore, one might naturally expect that the local gravitational stability criterion is different from (\ref{Q1}) and (\ref{Q2}). Finding the generalized version of the Toomre's criterion in modified theories of gravity is not a new idea. For example, Milgrom (1989) has found the local stability criterion in modified Newtonian dynamics (MOND). Also Toomre-like criterion in the context of metric $f(R)$ gravity has been calculated for both gaseous and stellar disks by Roshan \& Abbassi (2014b). Furthermore, this issue has been investigated in the context of some modified gravity theories which introduce a Yukawa like term in the gravitational force by Habibi et al (2014). In this paper (Habibi et al 2014) the local stability of disks in the context of MOG, has been also studied briefly as an special example for the above mentioned theories. Also some approximate stability criteria has been derived. Here, we give a more careful treatment and find the exact form of the stability criteria and use the relevant observational data to find out a detectable difference between Newtonian gravity and MOG.

\section{Modified gravity (MOG)}
In this section we briefly review the current status of MOG among other extended theories of gravity. Also, we briefly review its weak field limit which is necessary to achieve the main goal of this paper.

As we mentioned in the introduction, MOG is a fully relativistic and covariant generalization of GR. Moffat (2006) introduced this theory as an alternative to dark matter theories. MOG is much more complicated than GR in the sense that its associated gravitational fields are more than GR. In fact, MOG is a Scalar-tensor-vector theory of gravity, while GR is a tensor theory. In other words, in GR the components of the metric tensor are the gravitational fields of the theory. On the other hand, in MOG in addition to the metric tensor, there are three scalar fields ($\omega(x^{\beta}), \mu(x^{\beta})$ and $G(x^{\beta})$) and also a massive Proca vector field $\phi^{\beta}$ ( in the current literature of MOG, the scalar field $\omega$ is assumed to be constant). Therefor, MOG has equipped with some degrees of freedom in order to handle the dark matter problem without invoking dark matter particles. This theory has been successfully applied to explain the rotation curves of spiral galaxies and the mass discrepancy in the galaxy clusters (Brownstein \& Moffat 2006, 2007; Moffat 2006; Brownstein 2009; Moffat \& Toth 2008, 2009, 2013; Moffat \& Rahvar 2013, Moffat \& Rahvar 2014).

For the details of deriving the weak field limit of MOG, we refer the reader to Moffat \& Rahvar (2013), Roshan \& Abbassi (2014a). In the weak field limit of this theory, the particle's equation of motion can be written as
\begin{equation}
\frac{d^2\mathbf{r}}{dt^2}=-\nabla \Phi
\end{equation}
where $\Phi$ is a effective gravitational potential and is given by 
\begin{equation}
\Phi=\Psi+ \chi \phi^{0}
\end{equation}
where $\chi$ is a coupling constant and $\phi^0$ is the zeroth component (or time component) of the vector field. Furthermore, $\Psi$ and $\phi^0$ satisfy the following equations
\begin{equation}
\nabla^2\Psi=4\pi (1+\alpha) G \rho
\label{var}
\end{equation}
\begin{equation}
(\nabla^2-\mu_0^2)\chi \phi^0=-4\pi\alpha G\rho
\label{vec}
\end{equation}
where $\mu_0$ and $\alpha$ are the free parameters of the theory in the weak filed limit, $G$ and $\rho$ are the gravitational constant and the matter density, respectively. In fact, $\mu_0$ is the background value of the scalar field $\mu$ and $\alpha$ is related to the coupling constant $\chi$ as $\alpha=\chi^2/\omega_0G$. Note that $\omega_0$ is the background value of the scalar field $\omega$. For moe details see Roshan \& Abbassi (2014a). The observational values of the free parameters $\alpha$ and $\mu_0$ are known from rotation curve data of spiral galaxies. It has been shown by Moffat \& Rahvar (2013) that the best values for these parameters are $\alpha=8.89\pm 0.34$ and $\mu_0=0.042\pm 0.004 kpc^{-1}$. It is worth mentioning that, equations (\ref{var}) and (\ref{vec}) can be combined to write the generalized Poisson equation as 
\begin{equation}
\nabla^2\Phi=4\pi G\rho+\alpha\mu_0^2G\int\frac{e^{-\mu_0|\mathbf{r}-\mathbf{r}'|}}{|\mathbf{r}-\mathbf{r}'|}\rho(\mathbf{r}')d^3x'
\label{gp}
\end{equation}
For considering the dynamics of a self-gravitating fluid system in the context of MOG, in addition to the generalized Poisson equation (\ref{gp}), we need the generalized version of the continuity and Euler equations. It can be shown that the mathematical form of these equations are the same as in Newtonian gravity (Roshan \& Abbassi 2014a). The continuity and Euler equations are:
\begin{equation}
\frac{\partial \rho}{\partial t}+\nabla \cdot (\rho \mathbf{v})=0
\label{conti}
\end{equation}
\begin{equation}
\frac{\partial \mathbf{v}}{\partial t}+(\mathbf{v}\cdot \nabla)\mathbf{v}=-\frac{\nabla p}{p}-\nabla \Phi
\label{euler}
\end{equation}
where $p$ is the pressure and $\mathbf{v}$ is the velocity of the fluid. Equations (\ref{gp})- (\ref{euler}) combined with the equation of state of the fluid, makes a complete set of equations for describing the dynamics of a fluid system in the framework of MOG. Equivalently, one can use equations (\ref{var}) and (\ref{vec}) instead of (\ref{gp}).

On the other hand, in order to investigate the dynamics of a stellar system we need the Boltzmann equation and the generalized Poisson equation (\ref{gp}). The Boltzmann equation in MOG has been investigated in Roshan \& Abbassi (2014a), Moffat \& Rahvar (2014). Again, one can verify that the Boltzmann equation's mathematical form does not change . The main difference is that the effective potential $\Phi$ appears in the Boltzmann equation instead of the Newtonian gravitational potential $\Phi_N$. Therefore, the collisionless Boltzmann equation in MOG is
\begin{equation}
\frac{\partial f}{\partial t}+\mathbf{v}\cdot \nabla f-\nabla\Phi\cdot \frac{\partial f}{\partial\mathbf{v}}=0
\label{bol}
\end{equation}
where $f$ is the phase-space distribution function. In the section \ref{stellar}, we will use (\ref{bol}) for studying the local stability of a stellar disk.

\section{dispersion relation for a self-gravitating fluid disk in MOG}\label{fluid}
In this section we study the behavior of a tightly wound density wave propagating in the surface of a fluid disk in the context of MOG. First, we use the modified Poisson equation (\ref{gp}) in order to calculate the gravitational potential of a tightly wound surface density. Then, by linearizing the field equations, we find the relevant dispersion relation. We assume that the background disk is razor-thin and axisymmetric and its plane corresponds to the $x-y$ plane. Also, we assume that the disk has a barotropic equation of state as $p=K \Sigma^{\gamma}$, where $K$ and $\gamma$ are constant real parameters. The model assumes a non-rotating cylindrical system $(r,\varphi, z)$, such that $z$ axis coincide with the rotation axis of the disk and the angle $\varphi$ increases in the direction of rotation. In the cylindrical coordinate, by assuming that the fluid is axisymmetric,  continuity and two components of Euler equation can be written as: 
\begin{equation}
\frac{\partial\Sigma}{\partial t}+\frac{1}{r}\frac{\partial}{\partial r}\left(\Sigma r v_{r}\right)+\frac{1}{r}\frac{\partial}{\partial\varphi}\left(\Sigma v_{\varphi}\right)=0
\label{conti}
\end{equation}
\begin{equation}
\frac{\partial v_{r}}{\partial t}+v_{r} \frac{\partial v_{r}}{\partial r}+\frac{v_{\varphi}}{r} \frac{\partial v_{r}}{\partial \varphi}-\frac{v_{\varphi}^{2}}{r}=- \frac{\partial}{\partial r}\left(\Phi+h\right)\\
\label{eulerr}
\end{equation}
\begin{equation}
\frac{\partial v_{\varphi}}{\partial t}+v_{r} \frac{\partial v_{\varphi}}{\partial r}+\frac{v_{\varphi}}{r} \frac{\partial v_{\varphi}}{\partial \varphi}+\frac{v_{\varphi} v_{r}}{r}=- \frac{1}{r}\frac{\partial}{\partial\varphi}\left(\Phi+h\right)
\label{eulerphi}
\end{equation}
where $v_{r}$, $v_{\varphi}$ and  $\Sigma$ are the velocity components in the radial and azimuthal directions and the surface density, respectively.
Furthermore, $h$ is the specific enthalpy defined as $h= \int dp/\Sigma$. Now, we linearize equations (\ref{conti})-(\ref{eulerphi}) by assuming that: $\Sigma=\Sigma_{0}+\Sigma_{1}$, $v_{r}=v_{r0}+v_{r1}=v_{r1}$, $v_{\varphi}=v_{\varphi 0}+v_{\varphi 1}$, $\Phi=\Phi_{0}+\Phi_{1}$ and $h=h_{0}+h_{1}$. The subscript "0" refers to the background value of the given quantity and "1" refers to the corresponding perturbed quantity. After linearizing, the fluid equations (\ref{conti})-(\ref{eulerphi}) can be written as (Binney \& Tremaine 2008):
\begin{equation}
\frac{\partial \Sigma_{1}}{\partial t}+\frac{1}{r}\frac{\partial}{\partial r}\left(\Sigma_{0} r v_{r1}\right)+\Omega \frac{\partial \Sigma_{1}}{\partial \varphi}+\frac{\Sigma_{0}}{r} \frac{\partial v_{\varphi 1}}{\partial \varphi}=0
\label{contin2}
\end{equation}
\begin{equation}
\frac{\partial v_{r1}}{\partial t}+\Omega \frac{\partial v_{r1}}{\partial \varphi}-2\Omega v_{\varphi 1}=-\frac{\partial}{\partial r}\left(\Phi_{1}+h_{1}\right)
\label{eulerr2}
\end{equation}
\begin{equation}
\frac{\partial v_{\varphi 1}}{\partial t}+\Omega \frac{\partial v_{\varphi 1}}{\partial \varphi}+\frac{\kappa^{2}}{2\Omega} v_{r1}=-\frac{1}{r}\frac{\partial}{\partial \varphi}\left(\Phi_{1}+h_{1}\right)
\label{eulerphi2}
\end{equation}
where $\Omega(r)$ is the angular rotation rate and the epicyclic frequency $\kappa$ is defined as
\begin{equation}
\kappa(r)=\sqrt{r \frac{d\Omega^{2}}{dr}+4 \Omega^{2}}
\label{epip}
\end{equation}
Also, equations (\ref{var}) and (\ref{vec}) can be linearized as
\begin{equation}
\nabla^2\Psi_1=4\pi (1+\alpha) G \Sigma_1 \delta(z)
\label{var2}
\end{equation}
\begin{equation}
(\nabla^2-\mu_0^2)\chi \phi_1^0=-4\pi\alpha G\Sigma_1 \delta(z)
\label{vec2}
\end{equation}
where $\delta(z)$ is the Dirac delta function. We choose infinitesimal disturbances of the form $Q_1=Q_a e^{i(kr+m\varphi-\omega t)}$. Where $\omega$ is the oscillation frequency and $k=\frac{2\pi}{\lambda}$ is the wavenumber. For the local stability analysis we use the Wentzel-Kramers-Brillouin (WKB) approximation (or tight winding approximation) which requires that $\frac{k r}{m}\gg 1$ (Binney \& Tremaine 2008). This approximation allows us to neglect terms proportional to $1/r$ compared with the terms proportional to $k$. Employing this approximation, equations (\ref{contin2})-(\ref{eulerphi2}) can be combined to give
\begin{equation}
(m \Omega-\omega)\Sigma_a+k\Sigma_0 v_{ra}=0
\label{conti3}
\end{equation}
\begin{equation}
v_{ra}=\frac{(m\Omega-\omega)k(\Phi_a+h_a)}{\Delta}
\label{eulerr3}
\end{equation}
\begin{equation}
v_{\varphi a}=\frac{2i B }{\omega-m\Omega}v_{ra}
\label{eulerphi3}
\end{equation}
where $\Delta$ and $B$ (the Oort's constant of rotation) are
\begin{equation}
\Delta=\kappa^2-(m\Omega-\omega)^2
\end{equation}
\begin{equation}
B(r)=-\frac{1}{2}\left(\Omega+\frac{d(\Omega r)}{dr}\right)
\end{equation}
In order to complete the local stability analysis, let us find the gravitational potential of a tightly wound spiral perturbation, $\Sigma_1=\Sigma_a e^{i(kr+m\varphi-\omega t)}$ in the neighborhood of an arbitrary point $(r_0,\varphi_0)$. 
Locally, the WKB density perturbation can be considered as a plane wave with wavenumber $\mathbf{k} = k\mathbf{e}_r$, where $\mathbf{e}_r$ is the unit vector in the radial direction. Therefor, finding the potential of a WKB perturbation reduces to finding the gravitational potential of a plane density wave. In order to find the gravitational potential of this perturbation, without loss of generality, we choose the $x$ axis to be parallel to $\textbf{k}(r_0)$. Then we use equations (\ref{var2}) and (\ref{vec2}) to find the gravitational potential. To do so, we guess the solution to (\ref{vec2}) as
\begin{equation}
\phi_{1}^0(x,y,z,t)=\phi^0_a~ e^{i (k x-\omega t)-|\zeta z|}
\label{po3}
\end{equation}
where
$\phi^0_a$
and
$\zeta$
are arbitrary constants. The disk is razor-thin and so there is no matter outside ($z\neq 0$) the disk. Therefore, at $z\neq 0$ equation (\ref{var2}) can be written as $\nabla^{2}\phi^0_1=\mu_0^2\phi_1^0$. Substituting the potential (\ref{po3}) into this equation, one can easily show that 
$\zeta=\pm \sqrt{k^2+\mu_0^2}$. On the other hand, since matter is located at $z=0$, derivative of $\phi^0_1$ with respect to $z$ is not continuous on the disk. In order to fix the parameter $\zeta$, we integrate equation (\ref{var2}) with respect to $z$ in the interval $z=-\xi$ to $z=+\xi$, where $\xi$ is a positive constant,
and then let $\xi\rightarrow 0$. The result is
\begin{equation}
\phi_1^0=\frac{2\pi G \alpha}{\chi\sqrt{k^2+\mu_0^2}}\Sigma_a
\end{equation}
a similar procedure for equation (\ref{vec2}) gives
\begin{equation}
\varphi_a=-\frac{2\pi G (1+\alpha)}{|k|}\Sigma_a
\end{equation}
finally, the gravitational potential of a plane wave in the context of MOG is given by
\begin{equation}
\Phi_a=-\frac{2\pi G}{|k|}\Sigma_a \left(1+\alpha-\frac{|k|\alpha}{\sqrt{k^2+\mu_0^2}}\right)
\label{gp2}
\end{equation}
Equations (\ref{conti3})-(\ref{eulerphi3}) and (\ref{gp2}) combined with $h_a=v_s^2\Sigma_a/\Sigma_0$, form a closed set of linear equations. We solve these equations to obtain the dispersion relation. Using equations (\ref{eulerr3}) and (\ref{gp2}) we can find a purely algebraic expression between $v_{ra}$ and $\Sigma_a$ as:
\begin{equation}
v_{ra}=\frac{(m\Omega-\omega)k}{\Delta}\left(\frac{v_s^2}{\Sigma_0}-\frac{2\pi G}{|k|} [1+\alpha-\frac{|k|\alpha}{\sqrt{k^2+\mu_0^2}}]\right)\Sigma_a
\end{equation}
substituting this equation into (\ref{conti3}), and restricting ourselves to the axisymmetric ($m=0$) perturbations, we obtain the following dispersion relation
\begin{equation}
\omega^2=\kappa^2+k^2v_s^2-2\pi G \Sigma_0 |k|\left(1+\alpha-\frac{|k|\alpha}{\sqrt{k^2+\mu_0^2}}\right)
\label{dis1}
\end{equation}
We derive the local gravitational stability criterion by using this dispersion relation. As expected, in the limit $\mu_0\rightarrow 0$ or $\alpha\rightarrow 0$, the dispersion relation for WKB perturbations in Newtonian gravity is recovered.  Since all the quantities in the right-hand side (RHS) of (\ref{dis1}) are real quantities, then the disk is stable against local disturbances if $\omega^2>0$ and unstable if $\omega^2<0$.

In order to find the local stability criterion, we write the dispersion relation (\ref{dis1}) in a dimensionless form. To do so, we define the dimensionless wavenumber $X$, and the parameter $\beta$ as
\begin{equation}
X=\frac{k}{\mu_0}~~~~,~~~~\beta=\frac{\mu_0 v_s}{\kappa}
\end{equation}
therefore the dispersion relation can be rewritten in the dimensionless form
 \begin{equation}
\frac{\omega^2}{\kappa^2}=1+\beta^2 X^2-\frac{2\beta |X|}{Q_g}(1+\alpha)+\frac{2\beta}{Q_g}\frac{\alpha X^2}{\sqrt{1+X^2}}
 \label{dis2}
 \end{equation}
where $Q_g$ is the dimensionless Toomre's parameter given by equation (\ref{Q2}). Therefore, the stability criterion $\omega^2>0$ takes the following form
\begin{equation}
Q_g>\frac{2\beta}{1+\beta^2 X^2}\left((1+\alpha)|X|-\frac{\alpha X^2}{\sqrt{1+X^2}}\right)
\label{dis30}
\end{equation}
This is the exact criterion for the stability of the fluid disk against a perturbation mode with given wavenumber $X$. Once we have the observational values of $\alpha$ and $\mu_0$, and also the magnitude of the wavenumber $X$, then we can explicitly determine the required value of the Toomre's parameter for providing the stability at the given point. More specifically, we should know the magnitude of the ratio $v_s/\kappa$ at the given location. It is natural that if we find the maximum value of the RHS of (\ref{dis30}) for given $\alpha$ and $\beta$, then we will get a criterion which is independent of the value of the wavenumber. Since this can not be done analytically, we have listed the local stability criterion for various values of $\alpha$ and $\beta$ in Table~\ref{tab1}. It is worth remembering that, by fitting
MOG to observed rotation curves of galaxies, the free parameters $\alpha$ and $\mu_0$ have been determined to have the values:
$\alpha = 8.89\pm0.34$ and $\mu_0 = 0.042 \pm 0.004~kpc^{-1}$. So in the Table~\ref{tab1}, we have chosen $\alpha$ as $8.89\pm 0.34$. On the other hand, in order to choose rational values for $\beta$, we have estimated it in the solar neighborhood. In the solar neighborhood the epicycle frequency is $\kappa\sim 37~\text{km s}^{-1}\text{kpc}^{-1}$, and the radial velocity dispersion is $\sigma_r\sim38~\text{km s}^{-1}$. Therefore $\beta\sim0.04$. It is obvious from Table~\ref{tab1} that in the solar neighborhood there is no significant difference between the local stability criterion of MOG and Newtonian gravity. Also, it is clear that the difference between these theories considering the local stability will be significant in the locations where the ratio of the sound speed to the epicycle frequency is large. In fact, $v_s/\kappa$ is an important length scale for comparing MOG with Newtonian gravity in the disk galaxies. If this length scale satisfies the inequality $v_s/\kappa \gtrsim 2.4~kpc$, the deviation between the local stability criterion of MOG and Newtonian gravity will be large. In the end of this section, we will discuss the relevant observational data for this length scale.
\begin{deluxetable}{ccccc}
\tabletypesize{\scriptsize}
\tablecaption{Stability criterion for fluid and stellar disks, for different values of $\alpha$ and $\beta$. \label{tab1}}
\tablewidth{0pt}
\tablehead{\colhead{$\alpha$} & \colhead{$\beta$} & \colhead{FLUID DISK} & \colhead{STELLAR DISK} }
\startdata
9.23	& 0.02		& $Q_g>1.0019$	& $Q_s>1.0014$		\\ 
9.23	& 0.04		& $Q_g>1.0075$	& $Q_s>1.0074$		\\ 
9.23	& 0.2		& $Q_g>1.4271$	& $Q_s>1.3502$	 \\ 
9.23	& 0.4		& $Q_g>2.5928$	& $Q_s>2.4840$	 \\ 
8.55	& 0.02		& $Q_g>1.0017$	& $Q_s>1.0013$	 \\ 
8.55    & 0.04	    & $Q_g>1.0069$	& $Q_s>1.0068$	 \\ 
8.55	& 0.2		& $Q_g>1.3536$	& $Q_s>1.2826$		\\ 
8.55	& 0.4		& $Q_g>2.4459$  & $Q_s>2.3452$		\\ 
\enddata
\end{deluxetable}

\begin{figure}
\centerline{\includegraphics[width=9cm]{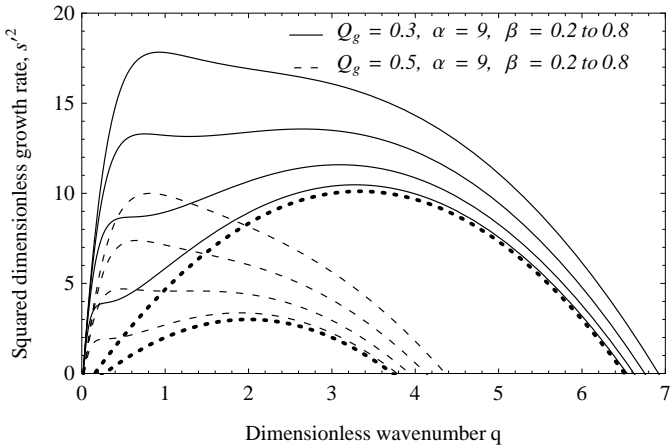}}
\caption[]{Solutions to the dispersion relation between the squared growth rate $s'^2$ and dimensionless wavenumber $q$ for a fluid disk. Solid curves correspond to $Q_g=0.3$ and the dashed curves are for $Q_g=0.5$. Higher value for parameter $\beta$ leads to greater instability at high wavelength.}
\label{fig1}
\end{figure}

It is interesting to plot the dispersion relation for the growing (unstable) modes and discuss the growing rates. Our purpose is to compare the growth rates in MOG and Newtonian gravity. To do so, we rewrite the dispersion relation (\ref{dis1}) as follows 
\begin{equation}
s'^2=\frac{2|q|}{Q_g}\left(1+\alpha-\frac{\alpha|q|}{\sqrt{q^2+\beta^2}}\right)-(1+q^2)
\end{equation}
 where the dimensionless wavenumber $q$ is defined as $q=kv_s/\kappa$ , and $s'=i\omega/\kappa$. Figure~\ref{fig1} shows the fluid dispersion relation for various values of $Q_g$ and $\beta$. The dotted curves are for Newtonian gravity. The maximum growth rate in MOG is higher than that of Newtonian gravity, and also occurs at smaller wavenumber. As in Newtonian gravity (dotted curves), the growth rates increase with decreasing $Q_g$. It is also clear form Figure~\ref{fig1} that, in MOG the range of instability is stretched to slightly larger wavenumber. Furthermore, by increasing the parameter $\beta$ the range of instability and also the maximum growth rate increase. 
 
 It is worth mentioning that in metric $f(R)$ gravity, by increasing the corresponding parameter $\beta$, the growth rate decreases (Roshan \& Abbassi 2014b). In this theory the parameter $\beta$ can be defined as $\beta=m_0 v_s/\kappa$, where $m_0$ is one of the free parameters. It is surprising that, although both MOG and $f(R)$ gravity lead to a stronger gravitational force relative to Newtonian gravity, MOG increases the growth rate while $f(R)$ gravity decreases it.

\begin{figure}
\centerline{\includegraphics[width=8.5cm]{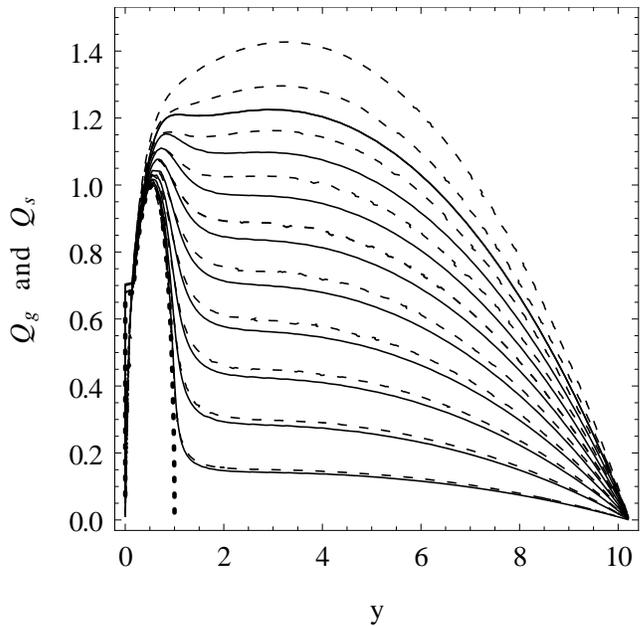}}
\caption[]{The boundary of stable and unstable axisymmetric perturbations in fluid and stellar disks. The chosen value for $\alpha$ is $9.23$ and $\beta$ varies from $0.02$ to $0.2$. The dotted line corresponds to Newtonian gravity and the dashed curves correspond to a fluid disk in MOG (different values of $\beta$). Furthermore, the solid lines show the neutral curves of a stellar disk.}
\label{fig2}
\end{figure}

It is also instructive to illustrate the neutral stability curves, i.e. $\omega=0$, for axisymmetric tightly wound spiral density waves in a fluid disk. These curves determine the boundary of the stable and unstable perturbations. Let us define a critical wavelength $\lambda_{\text{crit}}=2\pi/k_{\text{crit}}$ where $k_{\text{crit}}=\kappa^2/2\pi G\Sigma_0$. In Newtonian gravity $\lambda_{\text{crit}}$ is the largest unstable wavelength for a fluid disk with zero sound speed. Using this definition and the dispersion relation (\ref{dis1}), the line separating stable and unstable perturbations is the solution of the following equation
\begin{equation}
Q_g(y)=\sqrt{4y[1+\alpha-\frac{\alpha}{\sqrt{1+\left(\frac{2y\beta}{Q_g}\right)^2}}]-4y^2}
\label{bound}
\end{equation} 
where $y=\lambda/\lambda_{\text{crit}}$ and $\lambda=2\pi/k$. We have shown the neutral curves for various values of $\beta$ in Figure~\ref{fig2}. Furthermore, in this figure, we have assumed $\alpha=9.23$. The dotted curve corresponds to Newtonian gravity. Also, the dashed curves show the neutral curves of a fluid disk and solid lines are the corresponding lines for a stellar disk. It is clear that, at small wavelengths, the neutral curves of MOG and Newtonian gravity coincide. This coincidence is almost true for the interval $0<y<0.5$, i.e. $\lambda<0.5\lambda_{\text{crit}}$. However, for $y>0.5$ deviations between MOG and Newtonian gravity appear. More specifically, the range of instability is stretched to larger wavelength. It is surprising that the largest unstable wavelength in MOG is more than $10$ times the Newtonian case. In fact one can easily show that the largest unstable mode in MOG occurs at $\lambda=(1+\alpha)\lambda_{\text{crit}}$. This is a quiet big difference between these theories. However, one should note that at large wavelengths, there is a serious doubt for the validity of the WKB approximation. For example, for solar neighborhood $\lambda_{\text{crit}}\sim 10~kpc$. This wavelength is larger than the solar distance to the center of the disk galaxy, $r\sim 8.5~kpc$ . Therefore $k_{\text{crit}}r\sim 5.3$ and for the largest unstable wavelength in MOG we have $kr\sim 0.5$. Thus, obviously, the WKB approximation does not satisfied. Therefore, the wideness of the range of instability in MOG compared with the Newtonian gravity, can not be considered as a significant and detectable difference.

 On the other hand, it is clear in Figure~\ref{fig2} that increasing the parameter $\beta$, leads to a higher $Q_g$ for stability of all wavelengths. In other words, in MOG, higher $Q_g$ parameter compared with Newtonian gravity is needed to overcome the local gravitational instability. This result is also consistent with the Table~\ref{tab1}. As we discussed, this difference between MOG and Newtonian gravity could lead to a detectable deviation.  

The origin of this difference can be simply understood. Using the modified Poisson equation (\ref{gp}), the gravitational force between two point masses $m_1$ and $m_2$ that are located at $\mathbf{r_1}$ and $\mathbf{r}_2$ respectively, is given by
\begin{equation}
\mathbf{F}=\mathbf{F}_N+\alpha\mathbf{F}_N(1-e^{-\mu_0 r}(1+\mu_0 r))
\end{equation}
where $\mathbf{F}_N=-Gm_1m_2/r^2 \mathbf{e}_r$ is the Newtonian gravitational force and $\mathbf{r}=\mathbf{r_1}-\mathbf{r_2}$ and $\mathbf{e}_r$ is the unit vector in the direction of $\mathbf{r}$. Taking into account the current values of $\alpha$ and $\mu_0$, one can easily verify that $\mathbf{F}$ is stronger than Newton's gravitational force $\mathbf{F}_N$. Now, keeping in mind the physical meaning of the gravitational Jeans instability, it is natural to expect more gas pressure for preventing the gravitational collapse. In other words, it seems like that, when the gravitational force is stronger than the Newtonian gravity, then we need higher $Q_g$ parameter. However, this interpretation is not always true. For example, in the case of metric $f(R)$ gravity, where the corresponding gravitational force in the weak field limit is stronger than Newtonian gravity, the required value for $Q_g$ is smaller than the Newtonian one. 

\section{dispersion relation for a self-gravitating stellar disk in MOG}\label{stellar}
\begin{figure}
\centerline{\includegraphics[width=8cm]{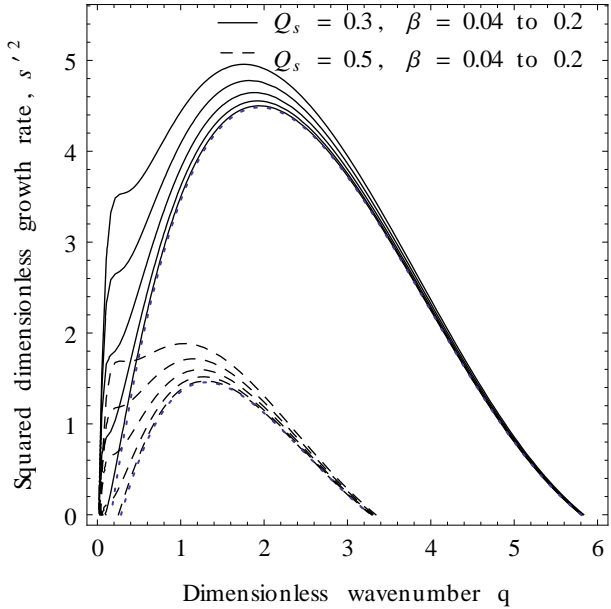}}
\caption[]{Solutions to the dispersion relation between the squared growth rate $s'^2$ and dimensionless wavenumber $q$ for a stellar disk. Solid curves correspond to $Q_g=0.3$ and the dashed curves are for $Q_g=0.5$. The dotted lines correspond to Newtonian gravity.}
\label{fig3}
\end{figure} 
Stability of stellar disk was first studied by Toomre (1964), Kalnajs (1965), and Lin \& Shu (1966) who pointed out that in certain regimes the stellar disk is stable against disturbances. When dealing with stellar disk stability, we need to use the collisionless Boltzmann equation (\ref{bol}), as well as the modified Poisson equation (\ref{gp}).  Since we assume that the stellar disk is collisionless, the stars can not create any pressure and consequently the sound speed for this system is zero. However, because of the epicycle motion at any given patch in the disk there are stars from different parts of the perturbed disk. This cause an important cancellation effect (Toomre 1964). In fact, it turns out that the stellar disk can be described by the Jeans equations, which their mathematical form is rather similar to the hydrodynamic equations (\ref{conti}) and (\ref{euler}). However, although equation (\ref{conti3}) does not change, equation (\ref{eulerr3}) is replaced with (see Roshan \& Abbassi (2014b) for more detail)
\begin{equation}
v_{ra}=\frac{m\Omega-\omega}{\Delta}k \Phi_a \mathcal{F}
\label{eulerr4}
\end{equation}
where $\mathcal{F}$ is the reduction factor given by (Binney \& Tremaine 2008)
\begin{equation}
\mathcal{F}(s,q^2)=\frac{1-s^2}{\sin\pi s}\int_0^{\pi}e^{-q^2(1+\cos\tau)}\sin s\tau \sin \tau d\tau
\end{equation}
where $s=-is'=(\omega-m\Omega)/\kappa$ and $q=k\sigma_r/\kappa$ and $\sigma_r$ are the dimensionless wavenumber and the radial velocity dispersion, respectively. Since the reduction factor does not depend on the form of Poisson equation (Roshan \& Abbassi 2014b), in MOG it possess the same general form and exhibit similar dependencies on dimensionless frequency and wave number as it has for Newtonian gravity. Now, substituting (\ref{eulerr4}) into equation (\ref{conti3}), we get the following dispersion relation
\begin{equation}
(\omega-m\Omega)^2=\kappa^2-2\pi G\Sigma_0 |k|[1+\alpha-\frac{\alpha|k|}{\sqrt{k^2+\mu_0^2}}]\mathcal{F}(s,q^2)
\label{dis3}
\end{equation}
at the limit $\alpha\rightarrow 0$ or $\mu_0\rightarrow 0$, the standard dispersion relation is recovered. Again, we restrict ourselves to axisymmetric perturbations. The system is stable against all axisymmetric modes if equation (\ref{dis3}) does not have a solution with negative $\omega^2$. In order to check this expectation, we rewrite dispersion relation (\ref{dis3}) as follows
\begin{equation}
1=\frac{2\pi|q|}{3.36Q_s}\left(1+\alpha-\frac{\alpha |q|}{\sqrt{q^2+\beta^2}}\right)\frac{\mathcal{F}(-is',q^2)}{1+s'^2}
\label{dis4}
\end{equation}
in this case $\beta=\mu_0 \sigma_r/\kappa$. It is straightforward to verify that the maximum value of the RHS of (\ref{dis4}) occurs at $s'=0$. Therefore, if the RHS of equation (\ref{dis4}) for $s'=0$ is smaller than $1$, then there is no perturbation with negative $\omega^2$ which satisfies the dispersion relation (\ref{dis4}). Consequently, the disk will be stable against all perturbations. Therefore, the stability criterion reads
 \begin{equation}
Q_s>\frac{2\pi|q|}{3.36}\left(1+\alpha-\frac{\alpha |q|}{\sqrt{q^2+\beta^2}}\right)\mathcal{F}(0,q^2)
 \label{dis5}
 \end{equation}
This criterion can be compared with the stability criterion for a fluid disk, i.e. equation (\ref{dis30}). The criterion (\ref{dis5}) can be replaced with a more simplified form
\begin{equation}
Q_s>\frac{2\pi}{3.36|q|}\left(1+\alpha-\frac{\alpha |q|}{\sqrt{q^2+\beta^2}}\right)\left(1-e^{-q^2}I_0(q^2)\right)
 \label{dis6}
 \end{equation}
 where $I_0$ is the modified Bessel function of order $0$. In deriving (\ref{dis6}), we have used the following identity
 \begin{equation}
 \mathcal{F}(0,q^2)=\frac{1}{q^2}\left(1-e^{-q^2}I_0(q^2)\right)
 \end{equation}
Equation (\ref{dis6}) is the main result of this section. For a given perturbation with wavenumber $q$, if we know the parameters $\alpha$ and $\beta$ then we will easily calculate the required value of $Q_s$ for stability of that perturbation. On the other hand, if we could find the maximum value of the RHS of equation (\ref{dis6}), then we will obtain a criterion which guarantees the stability against all wavelengths at every location on the disk. In Table~\ref{tab1}, the maximum value of the RHS have been calculated for various values of $\alpha$ and $\beta$. The tabulated data show that the required values of $Q_s$ for stabilizing the disk is larger than those for Newtonian case. As we mentioned before, this fact is also the case for a fluid disk.

In order to compare the growth rates in fluid and stellar disks, we have plotted the stellar disk dispersion relation (\ref{dis4}) in Figure~\ref{fig3}. This figure can be compared with Figure~\ref{fig1}. The solid curves correspond to $Q_s=0.3$ and $\beta=0.04$ to $0.2$. The dashed curves correspond to $Q_s=0.5$ and $\beta=0.04$ to $0.2$. Furthermore, the dotted lines are the corresponding Newtonian dispersion relations for the above mentioned values of $Q_s$. As in the case of a fluid disk, the growth rates increase with decreasing parameter $Q_s$.  Also larger $\beta$ leads to larger growth rate for large wavelengths. In fact, from Figure~\ref{fig3} it is evident that for small wavelengths there is no difference between the growth rates in MOG and Newtonian gravity. This point is also completely the same as for the fluid disk.

It is also instructive to plot the boundary between the stable and unstable modes. As we mentioned before, this curves can be specified by the requirement $\omega^2=0$. In this case, the dispersion relation (\ref{dis4}) can be rewritten as
\begin{equation}
\frac{f(Q_s,y)y}{1-e^{f(Q_s,y)}I_0(f(Q_s,y))}=\left[1+\alpha-\frac{\alpha}{\sqrt{1+\beta^2/f(Q_s,y)}}\right]
\label{dis7}
\end{equation}
where $f(Q_s,y)=\left(\frac{3.36Q_s}{2\pi y}\right)^2$. The curves corresponding to (\ref{dis7}) have been plotted in Figure~\ref{fig2} with solid lines for different values of $\beta$ ($0.02$ to $0.2$). It is obvious that similar to the fluid case, the range of instability is stretched to larger wavelength. The largest unstable wavelength is $\lambda=10.23 \lambda_{\text{crit}}$ (note that the we have chosen $\alpha=9.23$). However as we discussed in the previous section, this obvious deviation from the standard case does not necessarily lead to a significant difference between MOG and Newtonian gravity.
\section{discussion}
\begin{figure}
\centerline{\includegraphics[width=9cm]{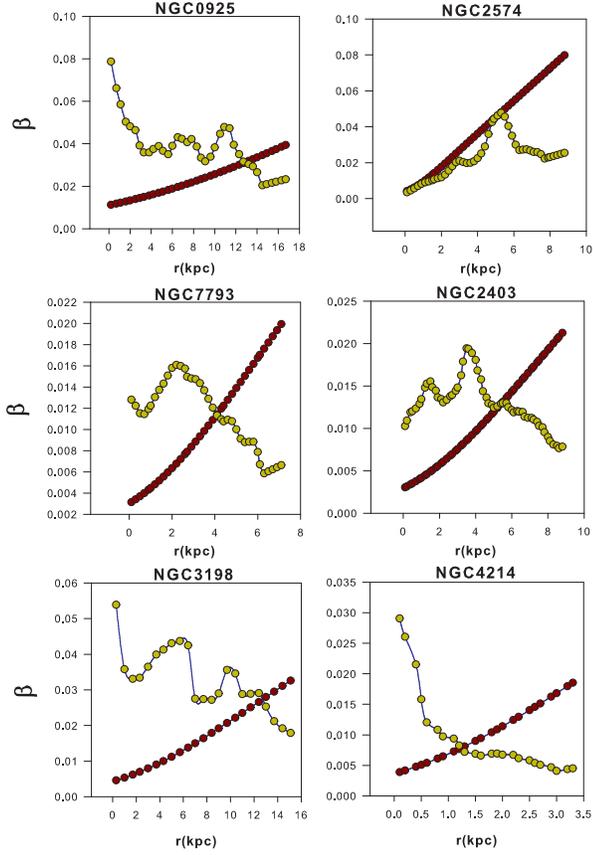}}
\caption[]{The parameter $\beta$ for stellar and gaseous components of some spiral galaxies. For gaseous component $\beta$ increases with radius. For the stellar component $\beta$ first increases and for large radius declines.}
\label{betathing}
\end{figure} 
\begin{figure}
\centerline{\includegraphics[width=9cm]{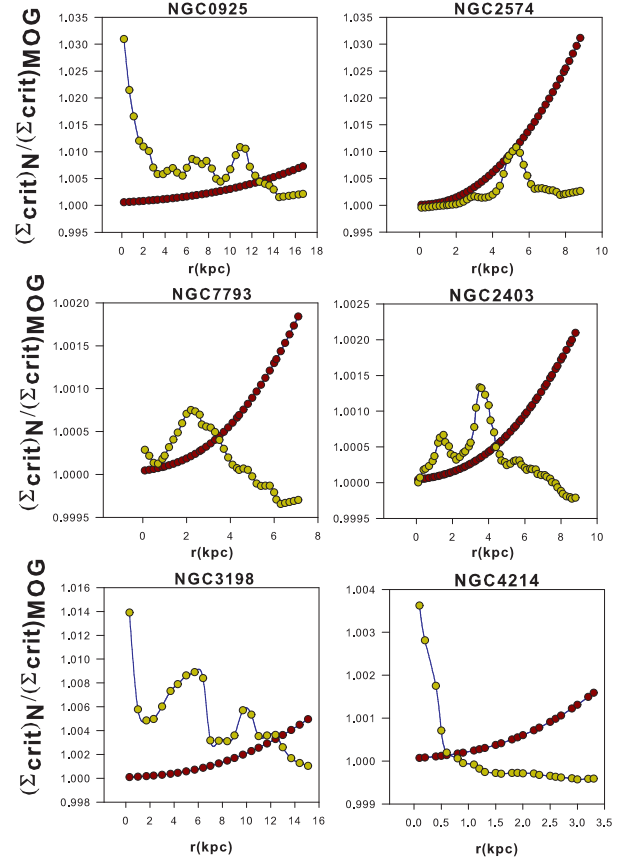}}
\caption[]{Ratio of the critical surface density of Newtonian gravity and MOG. The red circles shows the ratio for the stellar component and the green circles correspond to the gaseous component.}
\label{figcrit}
\end{figure} 
As we showed in sections \ref{fluid} and \ref{stellar}, the deviation between the stability criteria of Newtonian gravity and MOG is highly controlled by the magnitude of parameter $\beta$. Therefore, in order to compare these theories, it is essential to find the magnitude of this parameter with respect to radius from observational data in some real spiral galaxies. To do so, we use a sample of six galaxies of THINGS catalog of spiral galaxies (Leroy et al. 2008). We have chosen this sub-sample of galaxies just for illustration, and our results in this section are not restricted to these galaxies. For every galaxy in this catalog, the rotation curve $v(r)$, epicycle frequency $\kappa(r)$ and the velocity dispersion $\sigma_r (r)$ and $v_s$ are known. So we can easily calculate $\beta$ with respect to $r$ for both the stellar and gaseous component of the galaxies.

Before computing $\beta$ from observational data, it would be useful to make an order of magnitude estimation. It is worth mentioning that, at large radii in spiral galaxies the rotation curve is flat, i.e. $v(r)\sim v_{\text{flat}}$ where $v_{\text{flat}}$ is constant. Therefor, $\Omega(r)\sim \frac{v_{\text{flat}}}{r}$, and consequently $\kappa(r)=\sqrt{2}\frac{v_{\text{flat}}}{r}$. For gaseous component of the galaxies, $v_s$ is almost constant with respect to $r$ and is $v_s\sim 11~\text{km s}^{-1}$. In fact we have assumed that the sound speed is equal to the gas velocity dispersion $\sigma_{\text{gas}}$, see Leroy et al. (2008). Thus at large radii $\beta$ rises linearly with radius as
\begin{equation}
\beta(r)=\frac{\mu_0 v_s}{\sqrt{2}v_{\text{flat}}}r
\label{be}
\end{equation}
It means that the magnitude of $\beta$ is grater near the edge of the disk galaxy and consequently higher values for $Q_g$ is required to overcome the gravitational instability there. An immediate and important consequence follows from this equation. Near the edge of the galactic disk the difference between Newtonian gravity and MOG might be significant.  Therefor, the outer disks of spiral galaxies are exposed to instability more than the inner disks. Consequently, there might be some regions near the edge of the disk which can be stable in Newtonian gravity but unstable in the context of MOG. Such a deviation between these theories at large radii is expected because the free parameter $\mu_0^{-1}\sim 24 ~\text{kpc}$ is a large length scale compared to the characteristic size of the galaxies. A a consequence, taking into account the presence of the Yukawa term in this theory, we know that the effects of MOG appear at large radii. 

From Table~\ref{tab1}, it is evident that the stability criterion is significantly different from the standard case if $\beta\gtrsim 0.2$. Using equation (\ref{be}) the requirement $\beta\gtrsim 0.2$ reads
\begin{equation}
r\gtrsim 6.7 \left(\frac{v_{\text{flat}}}{v_s}\right) \text{kpc}
\end{equation}
If we set an average value $\frac{v_{\text{flat}}}{v_s}\sim 10$, then deviation between Newtonian dynamics and MOG will appear at $r\gtrsim 67~\text{kpc}$. On the other hand, almost for all spiral galaxies, the characteristic optical radius is smaller than $67~\text{kpc}$. Therefore, this quick estimation shows that it is unlikely to find a galaxy where the predictions of MOG are substantially different from that of Newtonian gravity.

However, let us do a more careful analysis by finding function $\beta(r)$ for our sample. The rotation curve $v(r)$ for the spiral galaxies of THINGS catalog can be written as (Boissier et al. 2003)
\begin{equation}
v(r)=v_{\text{flat}}\left[1-\exp\left(\frac{-r}{l_{\text{flat}}}\right)\right]
\label{n1}
\end{equation}
where $v_{\text{flat}}$ and $l_{\text{flat}}$ are free parameters that refer to the velocity at which the rotation curve is flat and the length scale over which it approaches this rotation speed, respectively. These parameters are not universal and are different from galaxy to galaxy. For these parameters we use Table~4 of Leroy et al. (2008). Also, using the rotation speed $v(r)$ the epicycle frequency can be written as
\begin{equation}
\kappa(r)=1.41\frac{v(r)}{r}\sqrt{1+b}
\label{n2}
\end{equation}
where $b=d \ln v(r)/d\ln r$. Furthermore, the radial velocity dispersion is given by (van der Kruit 1988; Leroy et al. 2008 )
\begin{equation}
\sigma_r(r)=1.55 \sqrt{Gl_{*}\Sigma_{*}(r)}
\label{n3}
\end{equation}
where $\Sigma_{*}$ is the stellar surface density and $l_{*}$ is the stellar scale length. We use Table 4 in Leroy et al. (2008) for $l_{*}$. For THINGS galaxies, $\Sigma_{*}$ has been measured with respect to radius in Leroy et al. (2008), for example see Table~7. Finally, using equations (\ref{n1})-(\ref{n3}), we have calculated $\beta$ for both the gaseous and stellar components of the sample. The result has been illustrated in Figure~\ref{betathing}. The red circles show $\beta(r)$ for gaseous component and green circles correspond to the stellar component. As we have already expected, for the gaseous component $\beta$ at large radii increases linearly with $r$. However, as illustrated in Figure~\ref{betathing}, although $\beta$ increases with radius, it remains small even at the outer disks. This happens also for the stellar component. In other words, $\beta$ remains small for stellar component. Of course, the behavior of $\beta$ with respect to $r$ for stellar component is completely different from that of the gaseous component. In fact, MOG predicts that at outer disks the gaseous component must be more unstable against local perturbations rather than the stellar disk. However, as we mentioned above, this difference is not too big to be significant and detectable.

In order to make the comparison between Newtonian gravity and MOG more clear, let us rewrite the stability criterion as follows
\begin{equation}
\Sigma(r)<\Sigma_{\text{crit}}(r)
\end{equation}
In other words, a local axisymmetric perturbation at $r$ is stable against gravitational collapse if the background matter density at that point is smaller than the critical density $\Sigma_{\text{crit}}(r)$. Where, $\Sigma_{\text{crit}}(r)$ in Newtonian gravity for gaseous and stellar disk are respectively
\begin{equation}
(\Sigma_{\text{crit}})_N=\frac{\kappa v_s}{\pi G}~~,~~(\Sigma_{\text{crit}})_N=\frac{\kappa \sigma_r}{3.36 G}
\end{equation}
On the other hand using criteria (\ref{dis5}) and (\ref{dis30}), one can write the critical surface density in MOG as follows
\begin{equation}
(\Sigma_{\text{crit}})_{\text{MOG}}=\frac{(\Sigma_{\text{crit}})_N}{H(\alpha,\beta)}
\end{equation}
Where $H>1$ and is a function of $\alpha$ and $\beta$. Note that for our purpose here, the exact form of the function $H$ is not needed. Therefore, we can conclude that the critical surface density in the context of MOG is smaller than the Newtonian gravity. This makes a disk more unstable than it would be in the standard case. In order to see the difference between $(\Sigma_{\text{crit}})_{\text{MOG}}$ and $(\Sigma_{\text{crit}})_N$, we have plotted their ratio in Figure~\ref{figcrit}. Although for the gaseous component the ratio $(\Sigma_{\text{crit}})_N/(\Sigma_{\text{crit}})_{\text{MOG}}$ increases linearly, but it does not get significantly larger than $1$. In fact this ratio can be approximated as
\begin{equation}
\frac{(\Sigma_{\text{crit}})_{\text{N}}}{(\Sigma_{\text{crit}})_{\text{MOG}}}\sim 1+ \beta
\end{equation}
Since $\beta$ remains small even at the outer disks, then one can conclude that the threshold matter density for gravitational collapse in Newtonian gravity and MOG is almost the same. Therefore, from this perspective, it seems like that there is no difference between these theories. It should be stressed that, we have done the same analysis for all galaxies in the THINGS catalog. And the result of this section is applicable to all of them.
\section{summary and conclusions}
In this paper, the local gravitational stability of galactic disks has been investigated in the context of MOG. The dispersion relation in the WKB approximation has been derived for both gaseous and stellar disks in the context of MOG. Using the relevant dispersion relations, we have found the local stability criteria for gaseous and stellar disks.


Finally, we have used a sample of spiral galaxies of the THINGS catalog in order to compare these theories using the current observational data. Our final results have been summarized in the Figures~\ref{betathing} and \ref{figcrit}. In Figure~\ref{figcrit}, we have compared the critical surface density over which the local perturbations collapse, in Newtonian gravity and MOG. For all galaxies of the sample, although the ratio of $(\Sigma_{\text{crit}})_N/(\Sigma_{\text{crit}})_{\text{MOG}}$ increases with radius, but it does not differ substantially form unity. Therefore, at least for all the spirals in the THINGS catalog, there is no significant difference between Newtonian gravity and MOG. It is also important noting that it could be more precise to find the stability criterion for a two-fluid disk (gas+stars). However, taking into account the main results of this paper, it is unlikely that a two-fluid analysis changes the results.

Of course, our results in this paper do not mean that the dynamics of spiral galaxies are the same in Newtonian gravity and MOG. Only the local perturbations fate is the same in these theories. It is worth remembering that MOG can explain the rotation curves of the spiral galaxies without using the dark matte halos. However, this can not be achieved in Newtonian gravity without invoking dark matter particles. Also, one should note that the global stability analysis of the disk galaxies in MOG, in principle, can be different from the standard case, for example see Brandao et al. (2010) for a numerical study of disk galaxies in MOG.

As a final remark, let us mention that there is a well-known instability in the dynamics of spiral galaxies called bar instability. In fact, in the early 1970s, the N-body simulations of self-gravitating disks showed that a cold rotationally supported disk is globally unstable to the formation of a non-axisymmetric system. In other words, a rotation dominated disk of particles will undergo a rapid transition to a bar-like pressure dominated system (for example see Hohl 1971). This result was puzzling because real galaxies like the Milky way does not show such a behavior. However, it is known that the presence of a dark matter halo can help to stabilize the disk, see Ostriker \& Peebles (1973). Furthermore, the presence of a dense bulge-like mass component near the center of the disk can also stabilize a cold self-gravitating disk, see Sellwood (2014) for a comprehensive review of the subject. Another approach to this problem can be provided via modified gravity theories. In alternative theories of dark matter there is no dark matter halo and the bar instability problem should be addressed without invoking dark matter. For example see Brada et al. (1999) and Tiret et al. (2007). In these papers the global stability of the disk galaxies has been investigated in MOND. As we mentioned before, MOG is an alternative theory of gravity presented for addressing the dark matter problem. Therefore, it would be interesting to study the bar instability problem in this theory. The result of the current paper can be used in numerical study of disk galaxies in MOG. More specifically, in order to set the initial conditions of the particles in a N-body simulation for studying the global stability of the disk, the local stability criterion on the disk is required to be satisfied.
\acknowledgments
We would like to thank the referee for useful comments. This work is supported by Ferdowsi University of Mashhad under Grant No. 100835 (25/05/1393).

\clearpage

\end{document}